\documentclass{elsarticle}
\biboptions{sort&compress}
\usepackage{hyperref}
\usepackage{threeparttable}
\usepackage{multirow}
\usepackage{longtable,booktabs}
\usepackage[]{caption2}

\journal{Physics Letters A}

\begin{document}

\begin{frontmatter}

\title{Collisional radiative model for the M1 transition spectrum of the highly-charged W$^{54+}$ ions}
\tnotetext[citation]{Please cite this article in press as: X.Ding et al., Collisional radiative model for the M1 transition spectrum of the highly-charged W$^{54+}$ ions, Phys. Lett. A (2018), \href{https://doi.org/10.1016/j.physleta.2018.05.046}{https://doi.org/10.1016/j.physleta.2018.05.046} .}


\author[nwnu,ustc]{Xiaobin Ding\corref{mycorrespondingauthor}}
\cortext[mycorrespondingauthor]{Corresponding author.}
\ead{dingxb@nwnu.edu.cn}
\author[nwnu]{Jiaoxia Yang}
\author[ustc]{Linfan Zhu}
\author[sophiaU]{Fumihiro Koike}
\author[NIFS,SOKENDAI]{Izumi Murakami}
\author[NIFS,SOKENDAI,KyushuU]{Daiji Kato}
\author[NIFS]{Hiroyuki A Sakaue}
\author[UEC]{Nobuyuki Nakamura}
\author[nwnu]{Chenzhong Dong\corref{mycorrespondingauthor}}
\ead{dongcz@nwnu.edu.cn}

\address[nwnu]{Key Laboratory of Atomic and Molecular Physics and Functional Materials
of Gansu Province, College of Physics and Electronic Engineering, Northwest Normal University, Lanzhou 730070, China}
\address[ustc]{Hefei National Laboratory for Physical Sciences at the Microscale and Department of Modern Physics, University of Science and Technology of China, Hefei, Anhui 230026, China}
\address[sophiaU]{Department of physics, Sophia University, Tokyo 102-8554, Japan}
\address[NIFS]{National Institute for Fusion Science, National Institutes of Natural Sciences, Toki, Gifu 509-5292, Japan}
\address[SOKENDAI]{Department of Fusion Science, SOKENDAI, Toki, Gifu, 509-5292,Japan}
\address[KyushuU]{Department of Advanced Energy Engineering Science, Kyushu University, Kasuga, Fukuoka, 816-8580, Japan}
\address[UEC]{Institute for Laser Science, The University of Electro-Communications, Chofu, Tokyo 182-8585, Japan}

\begin{abstract}
A detailed-level collisional-radiative model for the M1 transition spectrum of the Ca-like W$^{54+}$ ion as observed in an electron beam ion trap (EBIT) was constructed based on atomic data calculated by the relativistic configuration interaction method and distorted wave theory. The present calculated transition energy, rate and intensity of W$^{54+}$ M1 transitions are compared with previous theoretical and experimental values. The results are in reasonable agreement with the available experimental and theoretical data. The synthetic spectrum explained the EBIT spectrum in the 12-20 nm region, while a new possibly strong transition has been predicted to be observable with an appropriate electron beam energy. The present work provides accurate atomic data that may be used in plasma diagnostics applications.
\end{abstract}


\end{frontmatter}


\section{Introduction}

The structure and properties of tungsten ions draw more and more attention in recent research due to its application in magnetic confinement fusion. Tungsten has been chosen to be the material of the plasma-facing components in various fusion reactors, such as ITER, EAST, JET, and ASDEX-Upgrade, because of its favourable physical, chemical and thermal properties such as the ability to withstand high heat loads, low sputtering yield and tritium retention \cite{Matthews2009934,0029-5515-45-3-007}. Therefore, tungsten ions will inevitably be produced due to interaction between the edge plasma and the tungsten wall, and they will possibly be transported into the hot core plasma region. However, as tungsten is a heavy element with many electrons, it will not be fully ionized even in the core plasma with high temperature ($T\sim$ 25 keV) of ITER. Highly charged tungsten ions will emit high energy photons. Thus, large radiative energy loss from these ions can be expected, which will lead to disruption of the plasma if the relative concentration of W ions in the core plasma is greater than about 10$^{-5}$ \cite{Radtke2001}. Monitoring and controlling the flux of the highly charged W impurity ions will be crucial to the success of the fusion \cite{1402-4896-2009-T134-014022}. Furthermore, the spectra of W impurity ions observed from the fusion plasma provide important diagnostic tools to measure the electron density and temperature of the plasma.

Many research works on the radiative properties of tungsten ions related to fusion in various ionization stages have been published in the past several decades \cite{20072_S1060,Ding2012Collisional,Kramida2011,Kramida2009,Kramida2006,0953-4075-43-20-205004,Yanagibayashi2010,Fei2014,Murakami2015}. Special attention has been paid to the magnetic dipole (M1) transitions of the W$^{54+}$ ion \cite{Ralchenko2008,PhysRevA.83.032517,0953-4075-43-7-074026,0953-4075-44-19-195007,2015,Safronova2018,Zhao2018,1701-05504,Ding2017,0953-4075-50-4-045004}. In their electron beam ion trap (EBIT) experiment, Ralchenko \emph{et al.} \cite{Ralchenko2008} observed two emission peaks at 14.986 nm and 17.144 nm, which they identified with M1 transitions between the levels of the ground configuration [Ne]$3s^2 3p^6 3d^2$ of W$^{54+}$. They also observed another weaker transition peak of the same ion in the later work \cite{PhysRevA.83.032517}, where a non-Maxwellian collisional-radiative model was employed to analyse the observed spectrum. Safronova \emph{et al.} \cite{0953-4075-43-7-074026} calculated the rates of M1 transitions between the levels of the ground configuration by using the relativistic many-body perturbation theory (RMBPT). Quinet \cite{0953-4075-44-19-195007} calculated wavelengths and transition rates of forbidden transitions within the $3d^n$ ($n$=1-9) and $3p^k$ ($k$=1-5) ground configuration of tungsten ions (W$^{47+}$- W$^{61+}$) using fully relativistic multiconfiguration Dirac–Fock (MCDF). Guo \emph{et al.} \cite{2015} calculated the energy, transition wavelengths and rates for M1 transitions of W$^{54+}$ with the RMBPT and relativistic configuration interaction (RCI) methods. Recently, Safronova \emph{et al.} \cite{Safronova2018} and Zhao \emph{et al.} \cite{Zhao2018} calculated the M1 and E2 transition properties for the 3d$^n$ configuration in tungsten ions by using relativistic all-order many body calculation and MCDF methods, respectively. Ding \emph{et al.} \cite{1701-05504,Ding2017,0953-4075-50-4-045004} calculated the E1, E2, M1, M2 transitions data (the wavelengths, energy levels, radiative transition rates) of W$^{54+}$ by using the MCDF method with restricted active space. The results were in reasonable agreement with available experimental and theoretical values. They also constructed a detailed-level collisional-radiative model to simulate the visible spectrum of M1 transitions of W$^{26+} $ \cite{Ding2016Collisional} and the soft X-ray spectrum of E1 transitions of W$^{54+} $\cite{DING20187}, which had been observed in various EBIT experiments. The present work is devoted to interpretation of the observed M1 spectrum from the previous EBIT experiment \cite{PhysRevA.83.032517,Ralchenko2008}.

\section{Theoretical method}

Collisional-radiative modelling has been used to analyse the plasma spectrum. The emission intensity $I_{i,j}(\lambda)$ of an individual transition with wavelength $\lambda$ from the upper level $i$ to the lower level $j$ in an optically thin plasma is proportional to the transition energy $h\nu$, transition rate $A(i,j)$, population of the ions in the upper excited level $n(i)$ and the normalized line profile $\phi(\lambda)$, which can be written as:
\begin{equation}\label{eq1}
  I_{i,j}(\lambda) \propto  n(i) h\nu A(i,j) \phi(\lambda)
\end{equation}
where the normalized line profile $\phi(\lambda)$ was taken as a Gaussian profile, which may include the Doppler, natural, collisional and instrumental broadening effects. 

The population of the upper excited level $n(i)$ is determined by various atomic physics processes in the plasma, such as spontaneous radiative decay, electron-collisional (de)excitation, radiative recombination, electron-collisional ionization, dielectronic recombination, three-body recombination, and other interaction processes, e.g. between atom and ion or between ion and ion. However, the radiative recombination, three body recombination, and charge exchange processes are expected to be negligible in the EBIT transient dilute plasma. The free electron energy distribution in the electron beam of an EBIT facility is almost monoenergetic under typical operation conditions, the dielectronic recombination process can be neglected because the electron beam energy is far beyond the DR resonance energies. For the current interest, only one ionic species W$^{54+}$ are assumed to exist in the EBIT plasma, while electron-collisional ionization can be omitted. Finally, in the present calculation the excited state population is mostly determined by balance between the collisional (de)excitation and radiative decay processes.

The temporal development of the population $n(i)$ in a specific excited level \emph{i} can be obtained by solving the collisional-radiative rate equation:
\begin{eqnarray}
\frac{{\rm d}}{{\rm d} t}n(i)=&&\sum_{j<i}C(i,j)n_en(i) \nonumber \\
                              &&-[{\sum_{j<i}\left\lbrace F(i,j)n_e+A(i,j)\right\rbrace }+\sum_{j>i}C(i,j)n_e]n(i) \label{eq2}\\
                              &&+\sum_{j>i}[F(j,i)n_e+A(j,i)]n(j) \nonumber
\end{eqnarray}
where ${n_e}$ is the electron density of the plasma, $C(i,j)$ and $F(j,i)$ are collisional excitation and deexcitation rate coefficients from the level $j$ to $i$, respectively. The collisional excitation rate coefficient can be calculated by convoluting the cross section of the collisional excitation process with the free electron energy distribution function. For the plasma with the electron temperature $T_e$, the free electron energy distribution function was assumed to be Maxwellian. For the EBIT facility, the free electron energy distribution is almost monoenergetic and can be approximated by the $\delta$ function. The collisional deexcitation rate coefficients can be obtained by the detailed balance principle. These rate equations are solved under the quasi-steady-state (QSS) approximation, i.e., $ {\rm d}n(i)/{{\rm d} t} = 0$. The transition intensity $I_{i,j}(\lambda)$ can be calculated after the population $n(i)$ of the transition is determined.

W$^{54+}$ ion is a highly charged ion with the ground state [Ne]$3s^2 3p^6 3d^2$. It has two electrons in the outmost $3d$ subshell. To construct an appropriate CR model for the M1 spectrum of W$^{54+}$, we included all levels of the configurations formed by single and double electron substitution from the $n=3$ subshells of the ground configuration into the $3d$ and all $n=4$ subshells. There are 2,078 levels in the present CRM.  The configuration interaction effects were taken into account by the same scheme as in our previous work \cite{DING20187} which was found to adequately describe the most important electron correlation effects in the W$^{54+}$ ion. All the necessary fundamental atomic data, such as the energy levels, radiative transition rates (E1, M1, E2, M2), and cross sections of collisional (de)excitation were calculated by the RCI and distorted wave methods, as implemented in the Flexible Atomic Code (FAC) package \cite{Gu2008The}.

\section{Result and discussion}

The transition wavelength $\lambda$ (in nm), transition rate $A$ (in s$^{-1}$), and the intensity (emission power density of the plasma) "Int." of the M1 transitions among the $3s^2 3p^6 3d^2$ ground configuration levels of W$^{54+}$ are presented in Table~\ref{tab1}. ''$\lambda_{Exp}$'' and ''$\lambda_{Theo.}$'' represent the results from the EBIT experiment and other theories, such as MCDF, RMBPT, and RCI methods. It can be found from the table that the calculated wavelengths and transition rates are all in reasonable agreement with available experimental and other theoretical results. The discrepancy of the transition wavelengths between the present calculation and the EBIT experiment is within 0.34\%. In the previous MCDF calculation, Quinet \emph{et al.} \cite{0953-4075-44-19-195007} took the electron correlations within the $n = 3$ complex and some $n = 3$ to $n = 4$ single-excitation configurations into account. The discrepancies of the transition wavelengths and rates with the present calculation are about 0.07\% and 0.01\%, respectively. Ding \emph{et al.} \cite{0953-4075-50-4-045004} have included somewhat more electron correlations by extending the correlation configurations to include single and double excitation from the $n = 3$ complex to $n \le 6$ subshells, and the discrepancies of the transition wavelengths and rates with the present calculation are about 0.02\% and 0.03\%, respectively. Safronova \emph{et al.} \cite{0953-4075-43-7-074026} started their calculation based on the Dirac-Fock potential of the $1s^2 2s^2 2p^6 3s^2 3p^6$ core configuration for Ca-like tungsten ion by RMBPT. The discrepancies of the transition wavelengths and rates with the present calculation are about 0.24\% and 0.58\%, respectively. These large discrepancies are mainly due to the correlation effects of the $3s$ and $3p$ orbitals omitted in the RMBPT calculation. Ralchenko \emph{et al.} \cite{PhysRevA.83.032517} calculated the energy levels and M1 transition rates within the ground configuration of W$^{54+}$ by FAC. The configuration interaction within the $n = 3$ complex and single excitations up to $n = 5$ were taken into account in their calculation. Both the transition wavelength and rate agree very well with the present work, by about 0.02\% and 0.01\%, respectively. In order to ensure completeness, the configuration interactions within the $n = 3$ complex and the single and double excitations to $n = 4$ have been included more extensively in the present calculation. In Table~\ref{tab1}, we included only transitions with relatively large predicted intensity $I_{i,j}$, which could be observed in the experiment. Some of the transitions have large transition rates but small population resulting in small intensity that cannot be observed at the present electron beam energy and density of electrons. The uncertainties of the calculated transition rates are estimated by using the same method as in Zhao \emph{et al.} \cite{Zhao2018}. The MCDF transition rates from Ding \emph{et al.} \cite{0953-4075-50-4-045004} are taken as reference. Most of the strong transitions have uncertainties of less than 1 \%, while the weaker transitions have relatively larger uncertainties.  

The E2 transitions between levels of the ground configuration of W$^{54+}$ are included in the present calculation, but no significant contribution was found from these transitions. The E2 transition rates are on average 3 orders of magnitude smaller than M1 rates. For the level pairs having both M1 and E2 transitions allowed between them, the E2 contribution to the predicted intensity is about 1/1000 on average. For the level pairs allowing only E2 transition, there are no very large populations of the upper levels. Thus, the E2 transition contribution to the predicted intensities is also negligibly small. The M2 transitions are similar to E2 insofar as they do not give any significant contribution to the present model. 

The synthetic spectrum of the M1 transitions within the ground configuration of W$^{54+}$ in the wavelength range 12-20 nm is shown in Fig.~\ref{fig2}. The upper panel (a) is the spectrum obtained by convoluting the transition rates with a Gaussian profile. It shows the relative magnitude of the transition rates. The middle panel (b) is the synthetic spectrum for the EBIT case with the electron density $n_e = 10^{10}$ cm$^{-3}$ and the electron beam energy $E_e$ = 6000 eV. In the present calculation, all peaks observed in the EBIT experiment \cite{PhysRevA.83.032517} were reproduced, such as those corresponding to the transitions with key 4, 8 and 10. In addition to these three strong transitions, the peak 3 may be observable in future EBIT experiments under similar operating condition. This transition (key 3) is the M1 transition $(3/2,3/2)_2\rightarrow(3/2,5/2)_1$ within the ground configuration of the W$^{54+}$ ion with the predicted wavelength of 14.154 nm. The bottom panel (c) is the synthetic spectrum calculated for the plasma with the electron density $n_e = 10^{16}$ cm$^{-3}$ and electron temperature $T_e$ = 6000 eV with Maxwellian distribution of free electron energies. The result indicates that all transitions with large transition probabilities would be observable under these conditions.

The intensity for a specific transition is proportional to the population of the excited upper level $i$ and transition rate. Usually, a transition with a larger rate is more likely to be observed in an experiment. However, in the present case, the transition with key 1 has a large transition rate but was not observed in previous experiments, while another transition with key 4 was observed experimentally, but it has a small transition rate as shown in Fig~\ref{fig2} (a) and (b). As follows from our calculation, for the transition with key 4, the ratio of the population and depopulation fluxes is 11.69, while for the unobserved transition with the key 1 this ratio is only 0.107. This leads to a significant decrease of the population of the upper excited level of the transition with the key 1.

\begin{table}
\scriptsize
\begin{center}
\caption{The calculated transition wavelength $\lambda$, transition rate $A$ , intensity "Int.", available experimental and other theoretical values for strong M1 transitions of the W$^{54+}$ ion. The column 'Key' corresponds to the label in Fig~\ref{fig2}. The uncertainty "Unc." of the present calculation was estimated by using the method described in \cite{Zhao2018}.  Notion $a(b)$ for transition rates $A$ means a$\times$10$^b$.}\label{tab1}
\vspace{2mm}
\resizebox{\textwidth}{70mm}{
\begin{tabular}{p{1.3cm}p{1.3cm}p{0.8cm}p{0.9cm}p{1.9cm}p{0.8cm}p{2.1cm}p{1.5cm}p{0.5cm}p{0.5cm}}
\hline                                           
Lower & Upper & $\lambda$(nm) &	$\lambda_{Exp}$(nm)	& \centering{$\lambda_{Theo.}$(nm)}  &	$A$(s$^{-1}$)     & $A_{othe.}$(s$^{-1}$) & Int.(arb. u.)& Key & Unc.(\%)\\\hline
$(3/2,3/2)_2$	&$(5/2,5/2)_2$	&	7.703&	          & 7.712$^b$ 7.694$^c$ 7.700$^f$ 7.695$^g$ &	1.18(4)	&	1.28(4)$^b$1.15(4)$^c$ 1.14(4)$^f$ 1.14(4)$^g$ 	&	5.83E-11& & 0.6\\	
$(3/2,5/2)_1$	&$(5/2,5/2)_0$	&	12.720&	          & 12.721$^b$ 12.723$^c$ 12.757$^d$ 12.716$^f$ 12.738$^g$ &	7.85(6)	&7.32(6)$^b$7.88(6)$^c$ 7.83(6)$^d$7.79(6)$^f$ 7.86(6)$^g$&	9.91E+05&1 & 0.5\\	
$(3/2,5/2)_3$	&$(5/2,5/2)_2$	&	13.987&           &14.008$^b$  13.981$^c$ 13.997$^f$ 13.989$^g$ &	7.57(5)	&	7.52(5)$^b$7.59(5)$^c$ 7.48(5)$^f$7.58(5)$^g$	&	2.06E-09&2 & 0.5\\	
$(3/2,3/2)_2$	&$(3/2,5/2)_1$	&	14.154&	          &14.176$^b$  14.150$^c$ 14.163$^f$ 14.152$^g$	&	2.62(5)	&	2.58(5)$^b$2.63(5)$^c$ 2.60(5)$^f$2.63(5)$^g$  &	9.82E+07&3 & 0.5\\	
$(3/2,3/2)_2$	&$(3/2,5/2)_2$	&	14.986&14.959$^a$ &14.984$^a$  15.010$^b$  14.974$^c$ 14.980$^d$ (14.970,14.924)$^e$ 14.985$^f$ 14.977$^g$ &	1.81(6)	&	1.80(6)$^a$1.80(6)$^b$  1.82(6)$^c$1.81(6)$^d$	1.77(6)$^e$1.76(6)$^f$ 1.82(6)$^g$ &	5.30E+08 & 4  & 0.5\\		
$(3/2,5/2)_3$	&$(5/2,5/2)_4$	&	15.380	&	&15.413$^b$  15.369$^c$  15.369$^d$	15.378$^f$ 15.371$^g$ &	3.82(6)	&	3.76(6)$^b$3.84(6)$^c$	3.82(6)$^d$3.79(6)$^f$ 3.84(6)$^g$ &	1.13E+07 & 5 & 0.5\\	
$(3/2,5/2)_2$	&$(5/2,5/2)_2$	&	15.849	&	&15.860$^b$  15.827$^c$	15.848$^d$	15.839$^f$ 15.828$^g$ &	3.11(6)	&	3.10(6)$^b$3.13(6)$^c$	3.11(6)$^d$3.08(6)$^f$ 3.13(6)$^g$&	7.44E-09 & 6  & 0.6\\	
$(3/2,5/2)_1$	&$(5/2,5/2)_2$	&	16.899	&	&16.911$^b$  16.865$^c$	16.907$^d$ 16.874$^f$ 16.867$^g$  & 1.30(6)	&	1.29(6)$^b$1.31(6)$^c$	1.30(6)$^d$1.29(6)$^f$ 1.31(6)$^g$ &	2.93E-09 & 7 & 0.9\\	
$(3/2,3/2)_2$	&$(3/2,5/2)_3$	&	17.144	&	17.080$^a$&17.147$^a$  17.157$^b$ 17.110$^c$  17.138$^d$  (17.071,17.218)$^e$ 17.117$^f$ 17.105$^g$&	3.68(6)	&	3.68(6)$^a$3.68(6)$^b$  3.70(6)$^c$3.68(6)$^d$  3.64(6)$^e$3.66(6)$^f$ 3.70(6)$^g$ & 9.43E+08 & 8 & 0.5\\			
$(3/2,5/2)_4$	&$(5/2,5/2)_4$	&	18.607	&	&18.593$^b$  18.553$^c$  18.621$^d$	18.568$^f$ 18.550$^g$	&	1.10(6)	&	1.11(6)$^b$1.10(6)$^c$	1.09(6)$^d$1.09(6)$^f$ 1.10(6)$^g$ &	2.68E+06 & 9 & 0.5\\	
$(3/2,3/2)_0$	&$(3/2,5/2)_1$	&	19.268	&	19.177$^a$&19.281$^a$  19.294$^b$ 19.218$^c$	19.222$^d$  (19.160,19.422)$^e$ 19.237$^f$ 19.204$^g$&	1.72(6)	&	1.72(6)$^a$1.77(6)$^b$  1.74(6)$^c$1.74(6)$^d$  1.71(6)$^e$1.72(6)$^f$ 1.74(6)$^g$&  4.74E+08 & 10 & 1.0\\			
$(3/2,5/2)_3$	&$(3/2,5/2)_4$	&	88.692	&	&90.123$^b$ 89.579$^c$ 89.510$^f$ 89.698$^g$ &	8.78(3)	&	8.56(3)$^b$8.49(3)$^c$ 8.41(3)$^f$ 8.45(3)$^g$&	5.31E-01 & & 2.1\\	
$(3/2,5/2)_3$	&$(3/2,5/2)_2$	&	119.045	&	&119.974$^b$ 119.920$^c$ 120.325$^f$ 120.370$^g$		 &	4.52(3)	&	4.35(3)$^b$4.43(3)$^c$ 4.32(3)$^f$ 4.38(3)$^g$	&  1.65E+05 & & 2.1\\	
$(3/2,5/2)_2$	&$(3/2,5/2)_1$	&	255.047	&	&255.066$^b$ 255.047$^c$ 258.211$^f$ 257.030$^g$		 &	6.71(2)	&	6.79(2)$^b$6.57(2)$^c$ 6.40(2)$^f$ 6.56(2)$^g$	&  1.19E+04 & &3.2\\	

\hline
\end{tabular}}
\begin{tablenotes}
\item[] $^{\rm a}$ From Ralchenko et al. \cite{PhysRevA.83.032517}; Experimental data from EBIT observations and theoretical data from FAC calculation.
\item[] $^{\rm b}$ From Safronova et al. \cite{0953-4075-43-7-074026} by the RMBPT method.
\item[] $^{\rm c}$ From Ding et al. \cite{0953-4075-50-4-045004} by the MCDF method.
\item[] $^{\rm d}$ From Quinet \cite{0953-4075-44-19-195007} by the MCDF method.
\item[] $^{\rm e}$ From Guo et al. \cite{2015} by the RMBPT and RCI methods.
\item[] $^{\rm f}$ From Safronova et al. \cite{Safronova2018} by the relativistic all-order many body calculation methods.
\item[] $^{\rm g}$ From Zhao et al. \cite{Zhao2018} by the MCDF methods.
\end{tablenotes}
\end{center}
\end{table}

\begin{figure}
\label{fig2}
\centering
\includegraphics{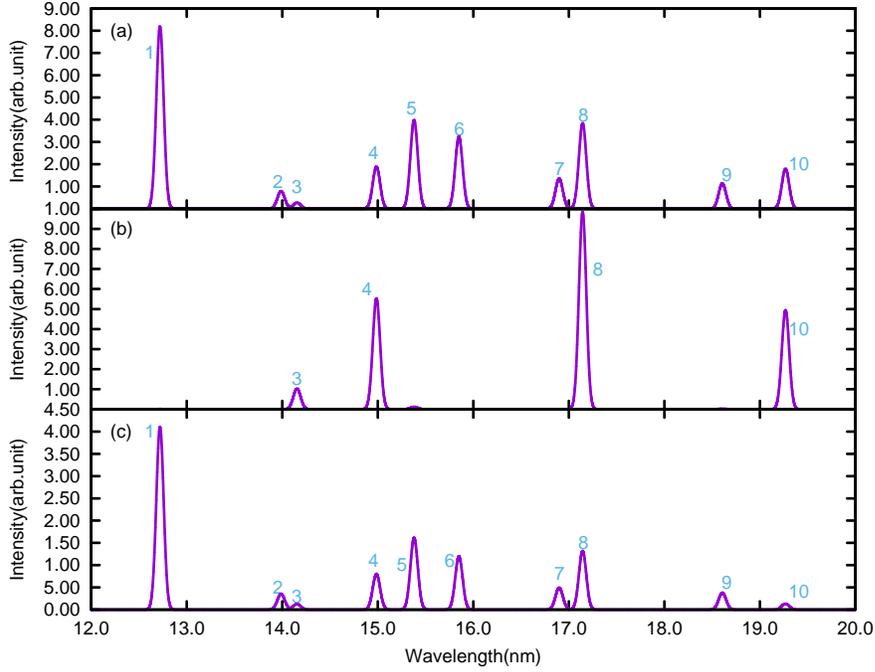}
\caption{A synthetic spectrum of the W$^{54+}$ ion for wavelengths 12-20 nm: (a) Convoluting the transition rate with a Gaussian profile; (b) The spectrum for the EBIT case with the electron density $n_e = 10^{10}$ cm$^{-3}$ and the electron beam energy $E_e$ = 6000 eV; (c) The spectrum for a plasma with the electron density $n_e = 10^{16}$ cm$^{-3}$ and temperature $T_e$ = 6000 eV.}\label{fig2}
\end{figure}

\section{Conclusion}

 A collisional-radiative model (CRM) of the M1 spectrum of the W$^{54+}$ ion previously observed in an EBIT experiment was constructed by considering the spontaneous radiative decay and electron-collisional (de)excitation processes in the plasma. The transition wavelengths and rates within the ground configuration [Ne]$3s^23p^63d^2$ of W$^{54+}$ were calculated using the relativistic configuration interaction method. All the necessary atomic data for constructing the CRM was calculated with the FAC package. The calculated energy levels and transition rates are in reasonable agreement with the previous theoretical and experimental values, and the synthetic spectrum provides a reasonable explanation for the EBIT experimental observation. Finally, a potentially strong M1 transition with a wavelength near 14.154 nm is predicted to be observable in future EBIT experiments.

\section*{Acknowledgment}

This work was supported by National Key Research and Development Program of China, Grant No. 2017YFA0402300, National Nature Science Foundation of China, Grant No. 11264035, Specialized Research Fund for the Doctoral Program of Higher Education (SRFDP), Grant No. 20126203120004, International Scientific and Technological Cooperative Project of Gansu Province of China (Grant No. 1104WCGA186), JSPS-NRF-NSFC A3 Foresight Program in the field of Plasma Physics (NSFC: No. 11261140328, NRF: 2012K2A2A6000443), the NINS program of Promoting Research by Networking among Institutions, Grant No. 01411702. Xiaobin Ding would like to thank the anonymous referee for their valuable comments and suggestions to improve the quality of the paper.  

\section*{References}
\footnotesize
\bibliography{abc}

\begin{thebibliography}{10}
\expandafter\ifx\csname url\endcsname\relax
  \def\url#1{\texttt{#1}}\fi
\expandafter\ifx\csname urlprefix\endcsname\relax\def\urlprefix{ }\fi
\expandafter\ifx\csname href\endcsname\relax
  \def\href#1#2{#2} \def\path#1{#1}\fi

\bibitem{Matthews2009934}
G.~Matthews, P.~Coad, H.~Greuner, M.~Hill, T.~Hirai, J.~Likonen, H.~Maier,
  M.~Mayer, R.~Neu, V.~Philipps, R.~Pitts, V.~Riccardo, Development of divertor
  tungsten coatings for the {JET ITER}-like wall, J. Nucl. Mater. 390-391
  (2009) 934--937.
\newblock \href {http://dx.doi.org/10.1016/j.jnucmat.2009.01.239}
  {\path{doi:10.1016/j.jnucmat.2009.01.239}}.

\bibitem{0029-5515-45-3-007}
R.~Neu, R.~Dux, A.~Kallenbach, T.~P\"{u}tterich, M.~Balden, J.~Fuchs,
  A.~Herrmann, C.~Maggi, M.~O'Mullane, R.Pugno, I.~Radivojevic, V.~Rohde,
  A.~Sips, W.~Suttrop, A.~Whiteford, the ASDEX Upgrade~team, Tungsten: an
  option for divertor and main chamber plasma facing components in future
  fusion devices, Nucl. Fusion 45 (2005) 209.
\newblock \href {http://dx.doi.org/10.1088/0029-5515/45/3/007}
  {\path{doi:10.1088/0029-5515/45/3/007}}.

\bibitem{Radtke2001}
R.~Radtke, C.~Biedermann, J.~L. Schwob, P.~Mandelbaum, R.~Doron, {Line and band
  emission from tungsten ions with charge $21+$ to $45+$ in the $45-70$
  {{\AA}}} \,range, Phys. Rev. A 64 (2001) 012720.
\newblock \href {http://dx.doi.org/10.1103/PhysRevA.64.012720}
  {\path{doi:10.1103/PhysRevA.64.012720}}.

\bibitem{1402-4896-2009-T134-014022}
C.~H. Skinner, Atomic physics in the quest for fusion energy and {ITER}, Phys.
  Scr. T134 (2009) 014022.
\newblock \href {http://dx.doi.org/10.1088/0031-8949/2009/T134/014022}
  {\path{doi:10.1088/0031-8949/2009/T134/014022}}.

\bibitem{20072_S1060}
M.~B. Chowdhuri, S.~Morita, M.~Goto, H.~Nishimura, K.~Nagai, S.~Fujioka, Line
  analysis of {EUV} spectra from molybdenum and tungsten injected with impurity
  pellets in {LHD}, Plasma Fusion Res. 2 (2007) S1060--S1060.
\newblock \href {http://dx.doi.org/10.1585/pfr.2.S1060}
  {\path{doi:10.1585/pfr.2.S1060}}.

\bibitem{Ding2012Collisional}
X.~Ding, I.~Murakami, D.~Kato, H.~A.~Sakaue, F.~Koike, C.~Dong,
  Collisional-radiative modeling of {W}$^{27+}$, Plasma Fusion Res. 7 (2012)
  2403128--2403128.
\newblock \href {http://dx.doi.org/10.1585/pfr.7.2403128}
  {\path{doi:10.1585/pfr.7.2403128}}.

\bibitem{Kramida2011}
A.~Kramida, Recent progress in spectroscopy of tungsten, Can. J. Phys. 89
  (2011) 551 -- 570.
\newblock \href {http://dx.doi.org/10.1139/p11-045}
  {\path{doi:10.1139/p11-045}}.

\bibitem{Kramida2009}
A.~Kramida, T.~Shirai, Energy levels and spectral lines of tungsten, {W III
  through W LXXIV}, At. Data Nucl. Data Tables 95 (2009) 305--474.
\newblock \href {http://dx.doi.org/10.1016/j.adt.2008.12.002}
  {\path{doi:10.1016/j.adt.2008.12.002}}.

\bibitem{Kramida2006}
A.~E. Kramida, J.~Reader, Ionization energies of tungsten ions: {W}$^{2+}$
  through {W}$^{71+}$, At. Data Nucl. Data Tables 92 (2006) 457--479.
\newblock \href {http://dx.doi.org/10.1016/j.adt.2006.03.002}
  {\path{doi:10.1016/j.adt.2006.03.002}}.

\bibitem{0953-4075-43-20-205004}
C.~S. Harte, C.~Suzuki, T.~Kato, H.~A. Sakaue, D.~Kato, K.~Sato, N.~Tamura,
  S.~Sudo, R.~D'Arcy, E.~Sokell, J.~White, G.~O'Sullivan, Tungsten spectra
  recorded at the {LHD} and comparison with calculations, J. Phys. B: At., Mol.
  Opt. Phys. 43 (2010) 205004.
\newblock \href {http://dx.doi.org/0953-4075/43/20/205004}
  {\path{doi:0953-4075/43/20/205004}}.

\bibitem{Yanagibayashi2010}
J.~Yanagibayashi, T.~Nakano, A.~Iwamae, H.~Kubo, M.~Hasuo, K.~Itami, {Highly
  charged tungsten spectra observed from JT-60U plasmas at T$_e$ $\approx$ 8
  and 14 keV}, Journal of Physics B: Atomic, Molecular and Optical Physics 43
  (2010) 144013.
\newblock \href {http://dx.doi.org/10.1088/0953-4075/43/14/144013}
  {\path{doi:10.1088/0953-4075/43/14/144013}}.

\bibitem{Fei2014}
Z.~Fei, W.~Li, J.~Grumer, Z.~Shi, R.~Zhao, T.~Brage, S.~Huldt, K.~Yao,
  R.~Hutton, Y.~Zou, Forbidden-line spectroscopy of the ground-state
  configuration of {Cd-like W}, Phys. Rev. A 90 (2014) 052517.
\newblock \href {http://dx.doi.org/10.1103/PhysRevA.90.052517}
  {\path{doi:10.1103/PhysRevA.90.052517}}.

\bibitem{Murakami2015}
I.~Murakami, H.~Sakaue, C.~Suzuki, D.~Kato, M.~Goto, N.~Tamura, S.~Sudo,
  S.~Morita, Development of quantitative atomic modeling for tungsten transport
  study using {LHD} plasma with tungsten pellet injection, Nuclear Fusion
  55~(9) (2015) 093016.
\newblock \href {http://dx.doi.org/10.1088/0029-5515/55/9/093016}
  {\path{doi:10.1088/0029-5515/55/9/093016}}.

\bibitem{Ralchenko2008}
Y.~Ralchenko, I.~N. Draganic, J.~N. Tan, J.~D. Gillaspy, J.~M. Pomeroy,
  J.~Reader, U.~Feldman, G.~E. Holland, {EUV} spectra of highly-charged ions
  {W}$^{54+}$-{W}$^{63+}$ relevant to {ITER} diagnostics, J Phys B: At , Mol
  Opt Phys 41 (2008) 021003.
\newblock \href {http://dx.doi.org/10.1088/0953-4075/41/2/021003}
  {\path{doi:10.1088/0953-4075/41/2/021003}}.

\bibitem{PhysRevA.83.032517}
Y.~Ralchenko, I.~N. Dragani\ifmmode~\acute{c}\else \'{c}\fi{}, D.~Osin, J.~D.
  Gillaspy, J.~Reader, Spectroscopy of diagnostically important magnetic-dipole
  lines in highly charged $3{d}^{n}$ ions of tungsten, Phys. Rev. A 83 (2011)
  032517.
\newblock \href {http://dx.doi.org/10.1103/PhysRevA.83.032517}
  {\path{doi:10.1103/PhysRevA.83.032517}}.

\bibitem{0953-4075-43-7-074026}
U.~I. Safronova, A.~S. Safronova, {Wavelengths and transition rates for nl-n'l'
  transitions in {Be-, B-, Mg-, Al-, Ca-, Zn-, Ag- and Yb-like} tungsten ions},
  J. Phys. B: At., Mol. Opt. Phys. 43 (2010) 074026.
\newblock \href {http://dx.doi.org/10.1088/0953-4075/43/7/074026}
  {\path{doi:10.1088/0953-4075/43/7/074026}}.

\bibitem{0953-4075-44-19-195007}
P.~Quinet, {Dirac-Fock} calculations of forbidden transitions within the 3p$^k$
  and 3d$^k$ ground configurations of highly charged tungsten ions
  ({W}$^{47+}$-{W}$^{61+}$), J. Phys. B: At., Mol. Opt. Phys. 44 (2011) 195007.
\newblock \href {http://dx.doi.org/10.1088/0953-4075/44/19/195007}
  {\path{doi:10.1088/0953-4075/44/19/195007}}.

\bibitem{2015}
X.~L. Guo, M.~Huang, J.~Yan, S.~Li, R.~Si, C.~Y. Li, C.~Y. Chen, Y.~S. Wang,
  Y.~M. Zou, {Relativistic many-body calculations on wavelengths and transition
  probabilities for forbidden transitions within the ground configurations in
  Co- through K-like ions of hafnium, tantalum, tungsten and gold}, J. Phys. B:
  At., Mol. Opt. Phys. 48 (2015) 144020.
\newblock \href {http://dx.doi.org/10.1088/0953-4075/48/14/144020}
  {\path{doi:10.1088/0953-4075/48/14/144020}}.

\bibitem{Safronova2018}
M.~S. Safronova, U.~I. Safronova, S.~G. Porsev, M.~G. Kozlov, Y.~Ralchenko,
  Relativistic all-order many-body calculation of energies, wavelengths, and
  {M1} and {E2} transition rates for the 3d$^n$ configurations in tungsten
  ions, Physical Review A 97~(1).
\newblock \href {http://dx.doi.org/10.1103/physreva.97.012502}
  {\path{doi:10.1103/physreva.97.012502}}.

\bibitem{Zhao2018}
Z.~Zhao, K.~Wang, S.~Li, R.~Si, C.~Chen, Z.~Chen, J.~Yan, Y.~Ralchenko,
  Multi-configuration {Dirac}{\textendash}{Hartree}{\textendash}{Fock}
  calculations of forbidden transitions within the 3d$^k$ ground configurations
  of highly charged ions ( {Z} = 72 {\textendash} 83 ), Atomic Data and Nuclear
  Data Tables 119 (2018) 314--353.
\newblock \href {http://dx.doi.org/10.1016/j.adt.2017.01.002}
  {\path{doi:10.1016/j.adt.2017.01.002}}.

\bibitem{1701-05504}
X.~Ding, S.~Rui, K.~Fumihiro, D.~Kato, I.~Murakami, H.~A. Sakaue, C.~Dong,
  Correlation, {Breit and Quantum Electrodynamics} effects on energy level and
  transition properties of {W}$^{54+}$ ion, Eur. Phys. J. D 71 (2017) 73.
\newblock \href {http://dx.doi.org/10.1140/epjd/e2017-70829-y}
  {\path{doi:10.1140/epjd/e2017-70829-y}}.

\bibitem{Ding2017}
X.~Ding, S.~Rui, K.~Fumihiro, I.~Murakami, D.~Kato, H.~A. Sakaue, N.~Nakamura,
  C.~Dong, Energy levels, lifetimes and radiative data of {W LV}, At. Data
  Nucl. Data Tables 119 (2018) 354--425.
\newblock \href {http://dx.doi.org/10.1016/j.adt.2017.02.002}
  {\path{doi:10.1016/j.adt.2017.02.002}}.

\bibitem{0953-4075-50-4-045004}
X.~Ding, S.~Rui, J.~Liu, K.~Fumihiro, I.~Murakami, D.~Kato, H.~A. Sakaue,
  N.~Nakamura, C.~Dong, {E1, M1, E2} transition energies and probabilities of
  {W}$^{54+}$ ions, J. Phys. B: At., Mol. Opt. Phys. 50 (2017) 045004.
\newblock \href {http://dx.doi.org/10.1088/1361-6455/aa53ec}
  {\path{doi:10.1088/1361-6455/aa53ec}}.

\bibitem{Ding2016Collisional}
X.~Ding, J.~Liu, F.~Koike, I.~Murakami, D.~Kato, H.~A. Sakaue, N.~Nakamura,
  C.~Dong, Collisional-radiative model for the visible spectrum of {W}$^{26+}$
  ions, Phys. Lett. A 380 (2016) 874--877.
\newblock \href {http://dx.doi.org/10.1016/j.physleta.2015.12.034}
  {\path{doi:10.1016/j.physleta.2015.12.034}}.

\bibitem{DING20187}
X.~Ding, J.~Yang, F.~Koike, I.~Murakami, D.~Kato, H.~A. Sakaue, N.~Nakamura,
  C.~Dong,
  \href{http://www.sciencedirect.com/science/article/pii/S0022407317304880}{{Theoretical
  investigation on the soft X-ray spectrum of the highly-charged W$^{54+}$
  ions}}, Journal of Quantitative Spectroscopy and Radiative Transfer 204
  (2018) 7 -- 11.
\newblock \href {http://dx.doi.org/https://doi.org/10.1016/j.jqsrt.2017.08.020}
  {\path{doi:https://doi.org/10.1016/j.jqsrt.2017.08.020}}.

\bibitem{Gu2008The}
M.~F. Gu, {The Flexible Atomic Code}, Can. J. Phys. 730 (2008) 127--136.
\newblock \href {http://dx.doi.org/10.1139/P07-197}
  {\path{doi:10.1139/P07-197}}.

\end{thebibliography}

\end{document}